\documentclass[11pt]{article} 
 
\usepackage{amssymb,cite} 
\usepackage{graphicx} 
\usepackage{latexsym} 

\usepackage{epsfig,amsfonts} 

\usepackage{bm}
\usepackage[ansinew]{inputenc}
\usepackage{rotating}
\usepackage{euscript}
 
\def\thefootnote{\fnsymbol{footnote}}

\renewcommand{\Re}{\, \mathrm{Re}\, }
\renewcommand{\Im}{\, \mathrm{Im}\, }

\newcommand{\pathC}{\mathcal C}

\newcommand{\fd}[1]{\left[\mathcal{D}#1\right]}
 
\newcommand{\eq}{\begin{equation}} 
\newcommand{\en}{\end{equation}} 
\newcommand{\be}{\begin{equation}} 
\newcommand{\ee}{\end{equation}} 
\newcommand{\eqa}{\begin{eqnarray}} 
\newcommand{\ena}{\end{eqnarray}} 
\newcommand{\ba}{\begin{eqnarray}} 
\newcommand{\ea}{\end{eqnarray}}

\newcommand{\br}{\langle} 
 
\newcommand{\bra}{\left \langle}
\newcommand{\ket}{\right \rangle}
\newcommand{\kt}{\rangle}

\newcommand{\um}{\frac12} 

\newcommand{\ZZ}{\hbox{{\rm Z{\hbox to 3pt{\hss\rm Z}}}}} 
\def\de{\partial}

\newcommand{\EQ}{\begin{equation}} 
\newcommand{\EN}{\end{equation}} 
\newcommand{\bea}{\begin{eqnarray}} 
\newcommand{\eea}{\end{eqnarray}}

\newcommand{\myfrac}[2]{\frac{\quad\displaystyle #1\quad}{\quad\displaystyle #2\quad}}

\begin{document} 
\begin{titlepage} 
\vskip0.5cm 
\begin{flushright} 
\end{flushright} 
\vskip0.5cm 
\begin{center} 
{\Large\bf  On the linear increase of the flux tube thickness near the deconfinement transition.} 
\end{center} 
\vskip1.3cm 
 
\centerline{A. Allais and    
M. Caselle}    
    
 \vskip0.4cm    
 \centerline{\sl  $^a$ Dipartimento di Fisica    
 Teorica dell'Universit\`a di Torino and I.N.F.N.,}    
 \centerline{\sl via P.Giuria 1, I-10125 Torino, Italy}    
 \centerline{\sl    
e--mail: \hskip 1cm (allais)(caselle)@to.infn.it}    
 \vskip1cm    
    
\begin{abstract}    
We study the flux tube thickness of a generic Lattice Gauge Theory near  
the deconfining phase transition.  It is well known that 
the effective string model predicts a logarithmic increase of the flux tube
thickness as a function of the interquark distance for any confining LGT at zero temperature. 
It is perhaps less known that this same model predicts a
linear increase in the vicinity of the deconfinement transition. We present a precise derivation of this result
and compare it with a set of high precision simulations in the case of the 3d gauge Ising model.
\end{abstract}    
\end{titlepage}    
 
\setcounter{footnote}{0} 
\def\thefootnote{\arabic{footnote}} 
\section{Introduction} 
\label{introsect}

The distinctive feature of the interquark potential in a confining gauge theory is that the  
colour flux is confined into a thin flux tube, joining the quark-antiquark pair. The quantum fluctuations of this flux tube can be well described by an effective string model. The most famous 
result of this model is the well known ``L\"uscher term'' which was predicted more than 25 years
ago~\cite{Luscher:1980fr,Luscher:1980ac} and was recently observed in high precision montecarlo simulations of lattice gauge theories (LGTs) both in (2+1) and in (3+1) dimensions with gauge groups ranging from  $Z_2$ to $SU(N)$~\cite{deForcrand:1984cz,Bali:1992ab, Luscher:2002qv, Juge:2002br, Bali:1994de,  Majumdar:2002mr, Caselle:2002ah, Caselle:2004er, Caselle:2007yc, Panero:2005iu, Lucini:2001nv, Gliozzi:2005ny, Athenodorou:2007du}.

Another well known prediction of the effective string theory is the  logarithmic increase of the width of the
 flux tube as a function of the interquark distance $R$. This behaviour was predicted many years ago by 
 L\"uscher, M\"unster and Weisz in~\cite{lmw80} and subsequently observed in various different models, 
 ranging again from SU(2) to $Z_2$ and to the pure gauge percolation model ~\cite{Bali:1994de,Pennanen:1997qm,
 cgmv95, Zach:1997yz, Koma:2003gi, Pfeuffer:2008mz, Giudice:2006hw}. Together with the linear rising
  of the interquark 
potential it has always been considered as one of the distinctive features of the confining regime in LGTs. 

A natural question is what happens of the flux tube width as the deconfinement temperature is approached  from
below.
According to the above
 picture one would naively expect that the log behaviour should hold in the whole confining phase.
 However we shall show in this paper that this is not the case. More precisely, we shall 
 show by means of high precision montecarlo simulations in the (2+1) dimensional $Z_2$ gauge model, that the 
 flux tube width also depends on the finite temperature of the theory 
 and that near the deconfinement temperature (but still in the confining phase) 
 the flux tube width increases linearly (and not logarithmically!) as a function of $R$.

As we shall see, this result is in perfect agreement with the effective string picture which indeed predicts 
a logarithmic increase at low temperature but shows a much more complex behaviour as the temperature increases
 and ultimately leads, as the deconfinement temperature is approached, to the linear behaviour 
observed in the simulations. 

This paper is organised as follows. In sect. 2 we define the flux tube thickness and discuss its evaluation in the framework of the effective string model both at zero and at finite temperature. In sect. 3 we present our montecarlo simulation while in sect. 4 we discuss our results and compare them 
with the effective string predictions. Sect. 5 is devoted to a few concluding remarks.

\section{Effective string prediction for the flux tube thickness} 
 
\subsection{Definition of the flux tube thickness} 
The lattice operator which is commonly used to evaluate the flux at zero temperature through a plaquette $p$ of the lattice is:
\eq 
\bra\phi(p;W)\ket=\frac{\bra W~U_p\ket}{\bra W \ket}-\bra U_p\ket 
\label{flux} 
\en 
where $W$ denotes a Wilson loop while $U_p$ denotes the operator associated with the plaquette $p$.

In a finite temperature setting we must substitute the Wilson loop with a pair of Polyakov loops. The lattice operator becomes in this case:
\eq 
\bra\phi(p;P,P')\ket=\frac{\bra P P'^\dagger~U_p\ket}{\bra PP'^\dagger \ket}-\bra U_p\ket 
\label{flux2} 
\en 
where $P$, $P'$ are two Polyakov loops separated by $R$ lattice spacings.

Within this setting the flux $\bra\phi(p;P,P')\ket$ depends on the spacelike coordinates of the plaquette, on its
orientation, on the separation $R$ of the Polyakov loops and on the length $L$ of the lattice in the timelike
direction. It does not depend on the timelike coordinate of the plaquette. Different possible orientations of the
plaquette $p$ measure different components of the flux. In the following we shall neglect this dependence which
plays no role if one is interested in the $R$ dependence of the flux tube width. Furthermore, since we are mainly
interested in the tube width half way between the two quarks, we restrict the plaquette to lie on the symmetry 
(hyper)plane half way between the two Polyakov loops. Under these conditions we have:
\[
 \bra\phi(p;P,P')\ket=\bra\phi(\vec h;R,L)\ket
\]
where $\vec h$ denotes the displacement of $p$ from the $P$ $P'$ plane. Along each of the directions spanned by $\vec h$, the flux density shows a gaussian like shape (see for instance Fig. 2 of ~\cite{cgmv95}). The width of this gaussian  is the quantity which is usually denoted as ``flux tube thickness'':

\eq
w^2(R,L)=\frac{\sum_{\vec h} \vec h^2 \bra\phi(\vec h;R,L)\ket}{\sum_{\vec h} \bra\phi(\vec h;R,L)\ket}
\label{w1}
\en

This quantity only depends on 
the interquark distance $R$ and  on the lattice size in the compactified timelike 
direction $L$, i.e. on the inverse temperature of the model. 
By tuning $L$ we can thus study the flux tube thickness near the deconfinement transition.
 
\subsection{Effective string model for the interquark potential} 
 
The starting point of the effective string description of the 
interquark potential is to model the latter in terms of a string
partition function:

\eq
\label{prp0conazeff}
\bra P P'^\dagger \ket = \int \left[ \mathcal{D} h \right] e^{-S_{\mbox{\tiny{eff}}}} \equiv Z(R,L)~,
\en
where 
$S_{\mbox{\tiny{eff}}}$ denotes the effective action for the world sheet spanned by the string. 
In (\ref{prp0conazeff}), the functional integration is done over world sheet configurations 
which have fixed boundary conditions along the space-like direction, and periodic boundary conditions 
along the compactified, time-like direction (the Polyakov lines are the fixed boundary of the string 
world sheet).

The simplest and most natural string model is the Nambu-Goto 
one, which assumes the string action $S_{\mbox{\tiny{eff}}}$ to be proportional to the area spanned 
by the string world sheet:
\eq
S_{\mbox{\tiny{eff}}}=\sigma\int d\tau\int  d\varsigma\sqrt{g}\ \ ,\label{action}
\en
where $g$ is the determinant of the two--dimensional metric induced on
the world--sheet by the embedding in $\mathbb R^d$  
and $\sigma$ is the string tension,
which appears as a parameter of the effective model.
\par
Eq.~(\ref{action}) is invariant with respect to reparametrization and Weyl
transformations. The standard choice to deal with these symmetries is to choose the so called
 ``physical gauge'' (see \cite{alvarez81} for more details) in which $g$ becomes a function 
of the transverse displacements of the string world-sheet only. These displacements 
(which we shall denote in the following as $h^i(\varsigma,\tau$) 
are required to satisfy the boundary conditions 
relevant to the problem --- in the present case,  periodic b.c. in the compactified direction and 
Dirichlet b.c. along the interquark axis direction:
\be
h^i(\tau+L,\varsigma)=h^i(\tau,\varsigma); \hskip 1cm h^i(\tau,-R/2)=h^i(\tau,R/2)=0\ \ .
\en
As it is well known, this gauge choice is anomalous\footnote{Another way to understand this anomaly is to notice
that this gauge fixing implicitly assumes that the world-sheet surface is a single-valued function of 
$(\tau,\varsigma)$, i.e. overhangs, cuts, or disconnected parts are excluded.}: 
rotational symmetry is broken at the quantum level 
unless the model is defined in the critical dimension $d=26$. However, this anomaly is known to  vanish 
at large distances \cite{olesen}, and this suggests to use the ``physical gauge'' for
an IR, effective string description also for $d \neq 26$.

If we restrict our attention to the $d=2+1$ case a further simplification occurs since  only one transverse degree of freedom ($h$) survives and the cross-interactions terms among
different transverse degrees of freedom disappear. In the physical gauge, (\ref{action}) takes the form: 
\eq
S[h]=\sigma\int_{-L/2}^{L/2}d\tau\int_{-R/2}^{R/2} d\varsigma\sqrt{1+(\de_\tau h)^2+(\de_\varsigma h)^2} \;\;.
\label{squareroot}
\en

Since we expect this model to be correct in the large $R$ limit the standard way to deal with the square root
term in the action is to perform a large $R$ expansion
 in powers of the dimensionless quantity $(\sigma RL)^{-1}$. 

\eq
\label{SexpansionNLO}
S[h]\sim \sigma L R+\frac{\sigma}{2}  \int_{-L/2}^{L/2}d\tau\int_{-R/2}^{R/2} d\varsigma
\left(\nabla h \right)^2 + O\left( (\sigma L R )^{-1} \right)\;\;,
\en

The first term is the classical contribution, it simply gives the area term in the interquark potential and we
shall neglect it in the following. The second term is a standard gaussian action
while the  higher order
$O((\sigma RL )^{-1})$ contributions  encode string self-interactions.

If we neglect in the expansion the string self-interaction terms (the so called ``gaussian approximation''), 
then using standard results of 2d conformal field theory (CFT) the partition function of the effective string
model can be evaluated exactly leading to the well known L\"uscher term. With some more effort also higher order
terms (and in particular 
the quartic self-interaction term written above) can
be evaluated~\cite{df82}

The resulting predictions have been compared with Montecarlo simulations of Polyakov loop correlators
for different gauge theories in the last
few years, showing a very good agreement at large distances and an increasing disagreement as smaller distances
and/or higher temperatures (i.e. smaller values of $L$) were approached (see for instance~\cite{Caselle:2002ah})

\subsection{The effective string width in the gaussian approximation}

The effective string approach allows to compute the flux tube width according to the following definition:
\[
  w^2(x;R,L)=\myfrac
  {\int_{\pathC}\fd{h}h^i(t,x)h^i(t,x)e^{-S[h]}}
  {\int_{\pathC}\fd{h}e^{-S[h]}}
\label{w2}
\]
where the sum is intended (as in the previous section) over all the surfaces bordered by the to Polyakov loops. Since the b.c. are periodic in the timelike direction, there is no dependence on $t$, and on setting $x=0$, i.e. half way between the two Polyakov loops, we obtain the effective string prediction for the flux tube as defined in (\ref{w1}).

The action $S[h]$ should be in principle the whole Nambu-Goto action (\ref{squareroot}), but, as anticipated, we shall truncate it to its gaussian approximation (\ref{SexpansionNLO}). In this way the width becomes a correlator in a 2d free bosonic theory:
\[
  w^2(x;R,L)=\left\langle h^i(t,x)h^i(t,x)\right\rangle
  \label{w3}
\]

This correlator is singular and must be regularized. The most natural choice is
a point splitting regularization:
\[
  w^2(x;R,L)=\left\langle h^i(t,x)h^i(t+\epsilon,x+\epsilon)\right\rangle
\label{w4}
\]

In fact, in the original lattice description, the flux density is evaluated using a plaquette operator which has an
intrinsic size of the order of the lattice spacing. This ultraviolet scale is translated in the effective string
description in the $\epsilon$ parameter of the point splitting regularisation. 

The correlator in (\ref{w4}) can be evaluated exactly (see the appendix). One can then perform an expansion in
$\epsilon$ of the result. As expected the first term diverges logarithmically while the remaining is finite and
describes the dependence of the result on the modular parameter $L/R$ of the cylinder.

It is important at this point to distinguish the two regimes: low and high temperature.
\begin{itemize}
\item
At low temperature, i.e. in the regime in which $L>> R$ the flux tube width in the gaussian
approximation is given by:
\eq
\begin{array}{c}
\displaystyle
\sigma w^2(z)=-\frac{1}{2\pi} \log\frac{ \pi |\epsilon|}{2 R} + 
\frac{1}{2\pi} \log\left|\theta_2\left(\pi \Re z/R\right)/\theta_1'(0) \right|\\ \\
q=e^{-\pi L/2R} 
\end{array}
\label{ris1}
\en
where (as discussed in the appendix) we use a complex coordinate $z$ to describe the cylinder bordered by the two Polyakov loops (with $\Re z$ representing the spacelike direction and $\Im z$ the timelike one) which are fixed in the positions $\Re z= \pm R/2$.
Setting $R_c= \pi |\epsilon|/2$ we see that the dominant term is:
\eq
\sigma w^2(z)= \frac{1}{2\pi} \log\frac{R}{R_c}
\en
as we anticipated, while the next to leading correction in the $L>> R$ limit turns out to be:
\[
  \frac{1}{2\pi} \log\left|\cos\left(\frac{\pi \Re z}{R}\right)\right|
\]
which vanishes if we choose $z=0$. 

\item
In the opposite regime $L<<R$ (i.e. high T, but still in the confining phase) 
the flux tube width has a very different expression:

\eq
\begin{array}{c}
  \displaystyle
  \sigma w^2(z)=-\frac{1}{2\pi} \log\frac{\pi |\epsilon|}{L} + \frac{1}{2\pi} \log\left|\theta_4(2\pi i \Re
  z/L)/\theta_1'(0) \right| +
  \frac{(\Re z)^2}{LR}\\ \\
  q=e^{-2 \pi R/L}
\end{array}
\label{ris2}
\en

This can be obtained by a modular transformation of the previous result or by direct calculation (see the
appendix). In this case in the $R>>L$ limit the dominant term turns out to be proportional to $\log L$
instead of $\log R$ and the (linear) $R$ dependence only appears in the first subleading correction.
In fact setting $z=0$ and using the relation:
$$
\theta_1'(0)=\theta_2(0)\theta_3(0)\theta_4(0)
$$
we obtain:
\[
  \sigma w^2(z)=-\frac{1}{2\pi} \log\frac{\pi |\epsilon|}{L} -\frac{1}{2\pi}\log|\theta_2(0)\theta_3(0)| ~~.
\]
Expanding this expression in powers of $q=e^{-2 \pi R/L}$ and setting $L_c=\pi|\epsilon|$ we
find~\footnote{Notice a misprint in the analogous expression reported in~\cite{Caselle:2006wr} where 
the linear coefficient was erroneously quoted to be 1/6 instead of 1/4.}:

\eq
  \sigma w^2(z)=\frac{1}{2\pi} \log\frac{L}{L_c} +
  \frac{R}{4L}-\frac{1}{\pi}e^{-2\pi\frac{R}{L}}+\cdots
\label{risfinale}
\en
\end{itemize}

This is the major result of this section and we shall devote the next sections to a check of this prediction
with a set of high precision Montecarlo simulations.

\section{Montecarlo simulations} 

Testing the logarithmic growth of the flux tube width
with Montecarlo simulations is a very difficult task since it requires
to study very large Wilson loops (or Polyakov loop correlators) and 
to control the statistical errors induced by the ratio of expectation values of (\ref{flux2})

Both these problems can be solved if one studies abelian LGTs for which a duality transformation can be implemented. In particular, for three dimensional LGTs, the dual model turns out to be a spin model. As discussed in~\cite{cgmv95} in this case one can study Polyakov loop correlators of arbitrary size by simply frustrating the links (in the dual lattice) orthogonal to the surface bordered by the Polyakov loops. The expectation value of the energy operator (which is dual to the plaquette of the original gauge theory) in this environment directly corresponds to the ratio of expectation values of (\ref{flux2}) thus solving at the same time also the second problem mentioned above.

This strategy was adopted in~\cite{cgmv95} to study the thickness of flux tubes generated by  
Wilson loops in the 3d gauge Ising model finding a perfect agreement with the predictions of the gaussian
approximation discussed above. By choosing different couplings and different Wilson loop sizes the authors of
\cite{cgmv95} were able to test the log growth over a range of more than two orders of magnitude.

The present paper can be considered as a natural continuation of the above analysis in the case of the Polyakov 
loop geometry which, as mentioned above, allows to introduce into the analysis also the finite temperature scale
 and allows to study the crossover from a log to a linear growth of the flux
tube thickness.

We report here for completeness a few details on the gauge Ising model, on the algorithm that we used and on the
parameters that we used in our simulations.

\subsection{The gauge Ising model}
The $3D$ $\ZZ_2$ gauge 
model on a cubic lattice is defined through the partition function:
\eq
Z_{gauge}(\beta)=\sum_{\{\sigma_l=\pm1\}}\exp\left(-\beta S\right)~~~,
\en
where the action $S$ is a sum over all the plaquettes of the cubic lattice:
\eq
S=-\sum_{\Box}\sigma_\Box~~~,~~~
\sigma_\Box=\sigma_{l_1}\sigma_{l_2}\sigma_{l_3}\sigma_{l_4}~~~.
\en

This model can be translated into the usual $3D$ Ising 
model  by the  Kramers--Wannier duality transformation:
\bea
Z_{gauge}(\beta)~\propto~ Z_{spin}(\tilde\beta)&&\\
\tilde{\beta}=-\um\log\left[\tanh(\beta)\right]~~&&~~,
\eea
where $Z_{spin}$ is the partition function of the Ising model on the 
dual lattice:
\eq
Z_{spin}({\tilde\beta})=\sum_{s_i=\pm1}\exp(-\tilde\beta H(s))~~~,
\en
with:
\eq
H(s)=-\sum_{\br ij \kt}J_{\br ij \kt}s_is_j
\en
where $i$ and $j$ denote nodes of the dual lattice and the sum is 
extended to the links ${\br ij \kt}$ connecting  the nearest--neighbour 
sites. For the moment the couplings $J_{\br ij \kt}$ are all chosen 
equal to $+1$ .

Using the duality transformation it is possible to build up a one--to--one mapping of physical observables of the gauge system onto the corresponding spin quantities. For instance, the vacuum expectation value of Polyakov loops correlator 
can be expressed in terms of spin variables as follows. First, choose an arbitrary surface $\Sigma$ bounded  by the two Polyakov loops; then ``frustrate'' the links of the dual lattice intersecting $\Sigma$, {\it i.e.} take $J_{\br ij \kt}=-1$ whenever $\br ij\kt\cap\Sigma\not=\emptyset$. Let us denote with $H'(s)$ the Ising Hamiltonian with this choice of couplings: the new Ising partition function $Z_{spin}'({\tilde\beta})=\sum_{s_i=\pm1}\exp\left(-\tilde\beta H'(s)\right)$ describes a vacuum modified by the two Polyakov loops, which we shall call the P--vacuum. It is easy to see at this point that, thanks to duality we  can write the expectation value of the Polyakov loops correlator as:

\eq
\bra PP'^\dagger\ket
~=~{Z_{spin}'\over Z_{spin}}~=~
\bra\prod_{\br ij\kt\cap\Sigma\not=\emptyset}
\exp(-2{\tilde\beta}s_is_j)\ket_{spin}~~~,
\label{pol}
\en
where  the product is over all the dual links intersecting $\Sigma$.

Similarly it is easy to see that:
\eq 
\frac{<PP'^\dagger U_p>}{<PP'^\dagger>}~=~ \br \exp(-2{\tilde\beta}s_ks_l)\kt_P
\label{pol2} 
\en 
where $s_ks_l$ is the link dual to the plaquette $U_p(x)$ and $<~~>_P$ denotes a mean value in the P-vacuum.

In this way we can immediately obtain the flux density 
by simply looking at the mean value of the (dual of the) plaquette in the model 
in which  all the links dual to $\Sigma$ are frustrated (see fig~\ref{fig:simulation}).

More precisely we have:

\eq 
\bra\phi(p,P,P')\ket=\bra U_p\ket_P-\bra U_p\ket  
\label{flux4} 
\en 
\subsection{Simulation setting}
\begin{figure}[htpb]
  \centering
  \includegraphics[width=3 in]{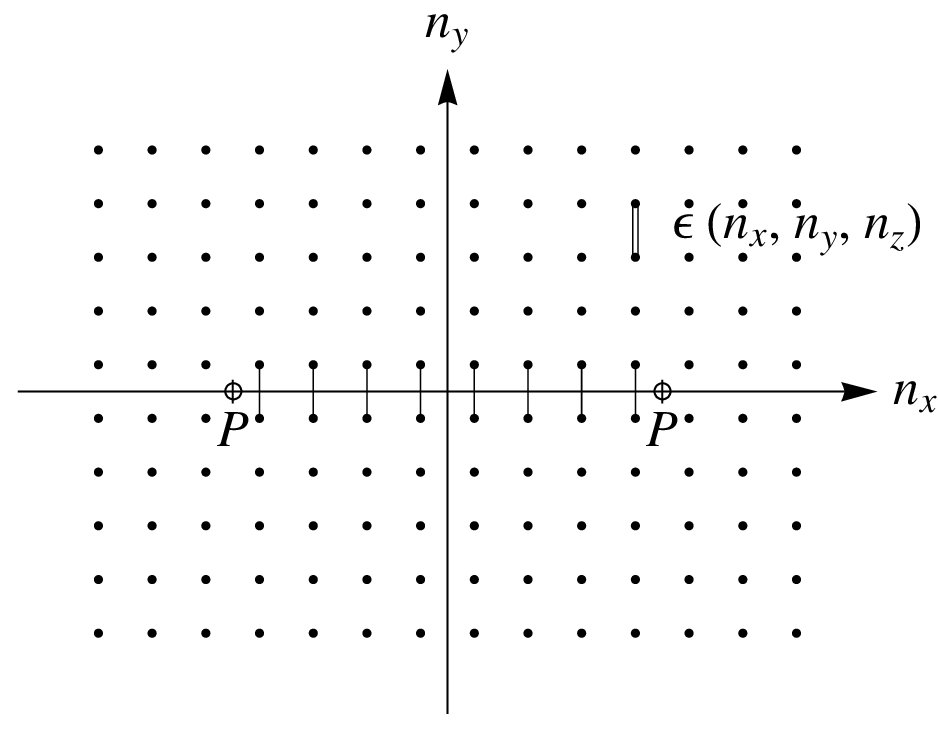}
  \caption{\small \textit{Schematic view of the simulation setting. The vertical bonds along the horizontal axis
   represent the frustrated links between the two Polyakov loops. The isolated bold face link represents the (dual
   of) plaquette operator.}} 
    \label{fig:simulation}
\end{figure}

We simulated the Ising model (both with and without frustrations) with a standard Swendsen-Wang algorithm.
We chose to measure the flux with a plaquette parallel to the surface joining the two Polyakov loops (i.e.
in the dual lattice, looking at the product of the spins joined by a link orthogonal to such surface, see
Fig.~\ref{fig:simulation}).
  We directly evaluated from the simulations the flux tube thickness and used a jackknife procedure to estimate
  the statistical errors. We performed all the simulations on a lattice of size $80\times80\times L$. We chose
   $\beta=0.75180$ for which the deconfinement
  transition is known to be located exactly at $L=8$~\cite{ch96}. Another reason for this choice is that for this
  value of $\beta$ the string tension is known with very high precision (see~\cite{Caselle:2007yc}) 
  $\sigma=0.0105255(11)$. We extracted the flux tube width for values
  of the interquark distance $R$ ranging~\footnote{Due to the finite 
  extent of the lattice size in the
 space directions ($L=80$) we expect finite size corrections to become important for $R>50$. In order to check
 this expectation we simulated  for one $L$ value: $L=10$ also a few values of $R$ larger than 50. }  
 from 5 to 50 and for $L\in\{9,10,11,12,14,16\}$, i.e. for values of the
  ratio $T/T_c$ ($T_c$ being the critical temperature) ranging from  $T/T_c=1/2$ to $T/T_c=8/9$. 
  
\section{Results} 

\begin{figure}[htpb]
  \centering
  \includegraphics[width=5 in]{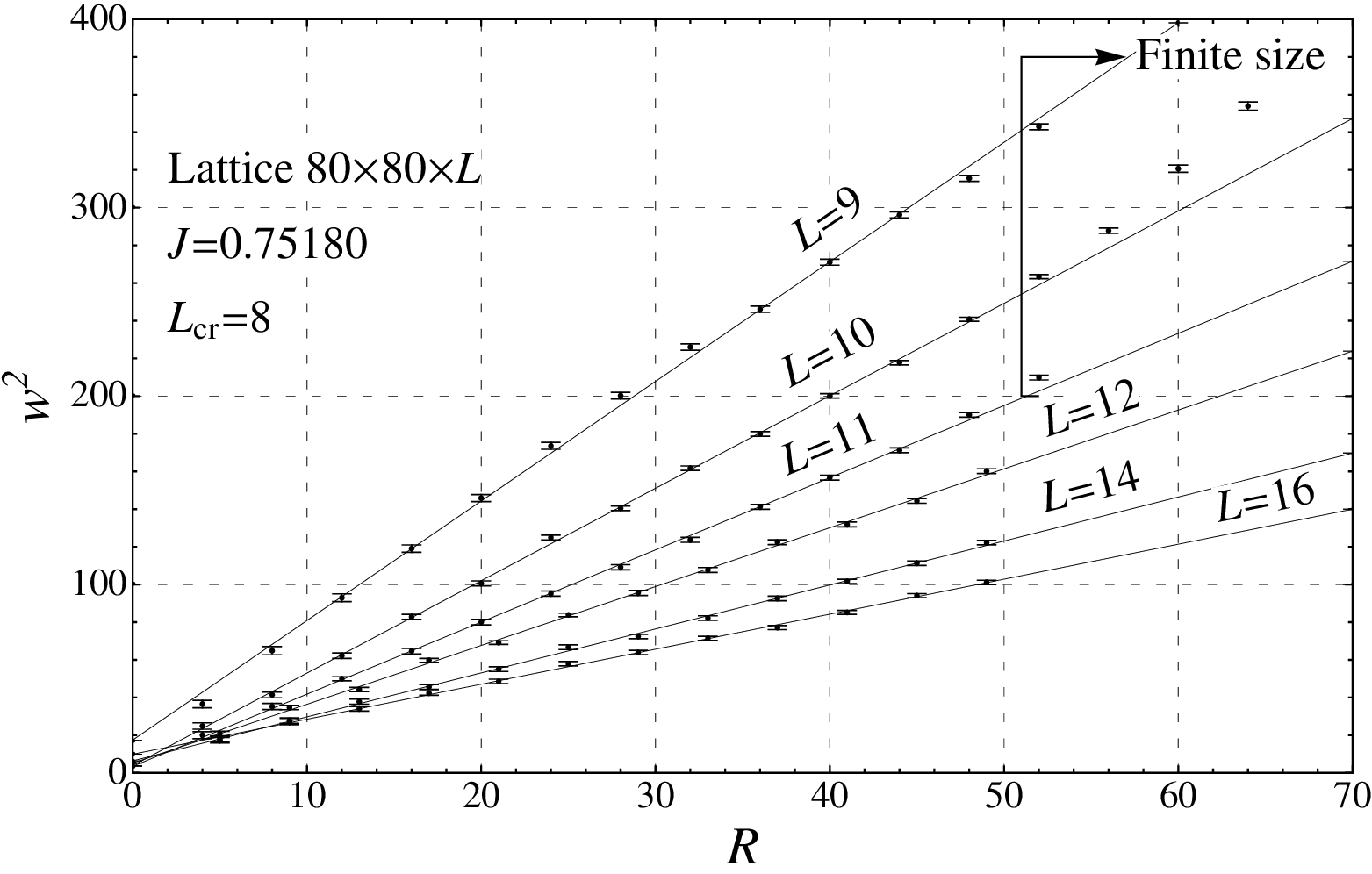}
  \caption{\small \textit{Flux tube thickness as a function of the interquark
  distance for various values of the inverse temperature $L$.}}
  \label{fig:simulation_results_08}
\end{figure}

We report  the results of our simulations in Tab. \ref{res1} and Tab. \ref{res2}. The same data are plotted in Fig. \ref{fig:simulation_results_08}. 

\begin{table}[htpb]
\centering
\begin{tabular}{|c|c|c||c|c|c||c|c|c|}
  \hline
 $L$&$R$ &  $w^2$     &$L$ &$R$ &  $w^2$     &$L$ &$R$ &  $w^2$     \\
  \hline\hline
  9 & 4  & 36.6(2.4)  & 10 & 4  & 24.9(2.0)  & 11 & 4  & 20.0(2.3)  \\ 
  9 & 8  & 64.8(2.6)  & 10 & 8  & 41.4(1.8)  & 11 & 8  & 35.2(2.0)  \\ 
  9 & 12 & 93.0(2.5)  & 10 & 12 & 62.2(1.7)  & 11 & 12 & 49.9(1.3)  \\ 
  9 & 16 & 119.0(2.4) & 10 & 16 & 82.8(1.6)  & 11 & 16 & 64.6(1.8)  \\ 
  9 & 20 & 145.8(2.3) & 10 & 20 & 100.6(1.6) & 11 & 20 & 80.0(1.7)  \\ 
  9 & 24 & 173.6(2.3) & 10 & 24 & 124.9(1.6) & 11 & 24 & 95.1(1.7)  \\ 
  9 & 28 & 200.3(2.0) & 10 & 28 & 140.4(1.4) & 11 & 28 & 109.1(1.7) \\ 
  9 & 32 & 226.0(2.0) & 10 & 32 & 161.7(1.4) & 11 & 32 & 123.7(1.6) \\ 
  9 & 36 & 246.1(2.0) & 10 & 36 & 179.9(1.4) & 11 & 36 & 141.2(1.6) \\ 
  9 & 40 & 271.0(1.9) & 10 & 40 & 200.1(1.3) & 11 & 40 & 156.7(1.6) \\ 
  9 & 44 & 296.2(1.9) & 10 & 44 & 217.7(1.3) & 11 & 44 & 171.2(1.6) \\ 
  9 & 48 & 315.6(1.9) & 10 & 48 & 240.8(1.3) & 11 & 48 & 190.0(1.4) \\ 
  9 & 52 & 342.9(1.8) & 10 & 52 & 263.4(1.3) & 11 & 52 & 209.8(1.4) \\ 
    &    &            & 10 & 56 & 287.8(1.7) &    &    &            \\
    &    &            & 10 & 60 & 320.7(2.3) &    &    &            \\
    &    &            & 10 & 64 & 353.9(2.6) &    &    &            \\
  \hline
\end{tabular}
\caption{\small \textit{Results of the simulations: the square width $w^2$ as a function of $R$ and $L$}}
\label{res1}
\end{table}
\begin{table}
\centering
\begin{tabular}{|c|c|c||c|c|c||c|c|c|}
  \hline
 $L$&$R$ &  $w^2$      &$L$ &$R$ &  $w^2$     &$L$ &$R$ &  $w^2$     \\
  \hline\hline
  12 & 5  & 20.9(1.6)  & 14 & 5  & 17.9(1.9)  & 16 & 5  & 17.3(1.8)  \\ 
  12 & 9  & 34.7(1.3)  & 14 & 9  & 27.7(1.7)  & 16 & 9  & 26.9(1.6)  \\ 
  12 & 13 & 44.3(1.3)  & 14 & 13 & 37.8(1.7)  & 16 & 13 & 34.0(1.6)  \\ 
  12 & 17 & 59.7(1.2)  & 14 & 17 & 45.6(1.6)  & 16 & 17 & 42.4(1.4)  \\ 
  12 & 21 & 69.1(1.2)  & 14 & 21 & 55.0(1.6)  & 16 & 21 & 48.6(1.4)  \\ 
  12 & 25 & 83.8(1.2)  & 14 & 25 & 66.7(1.4)  & 16 & 25 & 57.9(1.3)  \\ 
  12 & 29 & 95.5(1.6)  & 14 & 29 & 72.4(1.4)  & 16 & 29 & 63.9(1.3)  \\ 
  12 & 33 & 107.7(1.6) & 14 & 33 & 82.2(1.4)  & 16 & 33 & 71.5(1.3)  \\ 
  12 & 37 & 122.4(1.6) & 14 & 37 & 92.6(1.4)  & 16 & 37 & 77.1(1.3)  \\ 
  12 & 41 & 131.9(1.6) & 14 & 41 & 101.6(1.4) & 16 & 41 & 85.2(1.3)  \\ 
  12 & 45 & 144.4(1.6) & 14 & 45 & 111.3(1.4) & 16 & 45 & 94.1(1.3)  \\ 
  12 & 49 & 160.1(1.6) & 14 & 49 & 122.2(1.4) & 16 & 49 & 101.1(1.3) \\ 
  \hline
\end{tabular}
\caption{\small \textit{Same as Table \ref{res1}}}
\label{res2}
\end{table}

Looking at Fig. \ref{fig:simulation_results_08}, it is easy to see that, in agreement with our effective string calculations, the data for $L>9$ show a very nice linear behaviour as a function of $R$ in the range $10<R<50$. The data for $R>50$ show deviations due to finite size effects. 
We performed a linear fit  of the data in the range $10<R<50$ with the law:
\eq
  w^2=k(L)R+c(L);\quad k(L)=\frac{k_0}{\sigma L}
\en
where, according to the effective string calculation, we should have $k_0=1/4$
The results are reported in Tab. \ref{tab:simulation_fit_08}. 
  
\begin{table}[htpb]
\centering
\begin{tabular}{|c|c|c|c|c|c|}
  \hline
  $L$ & $k(L)$  & $c(L)$ & $k_0$ & $\sigma$               & $\chi^2_r$    \\
  \hline\hline
  9   & 6.19(10) & 23.0(3.2)  &  0.587(16)  &$4.48(7)\times10^{-3}$ & 2.6       \\
  10  & 4.90(4) & 4.1(1.5) &  0.516(5)  &$5.10(5)\times10^{-3}$ &  1.2     \\
  11  & 3.85(4) & 2.6(1.2) & 0.446(4)  & $5.90(6)\times10^{-3}$ &  0.8     \\
  12  & 3.14(4) & 4.4(1.2) &  0.397(5) & $6.63(8)\times10^{-3}$ &  1.1    \\
  14  & 2.33(3) & 6.3(1.2) &  0.344(5)  &$7.64(11)\times10^{-3}$ &  0.7  \\
  16  & 1.84(3) & 10.6(9) &   0.309(4) &$8.50(12)\times10^{-3}$&  0.5  \\
  \hline
\end{tabular}
\caption{\small \textit{Results of the fit  
$w^2=k(L)R+c(L)$ for various values of $L$. In the fifth column we also report the  values of 
the string tension $\sigma$ extracted from $k(L)$.}}
\label{tab:simulation_fit_08}
\end{table}

\begin{figure}[htpb]
  \centering
  \includegraphics[width=4 in]{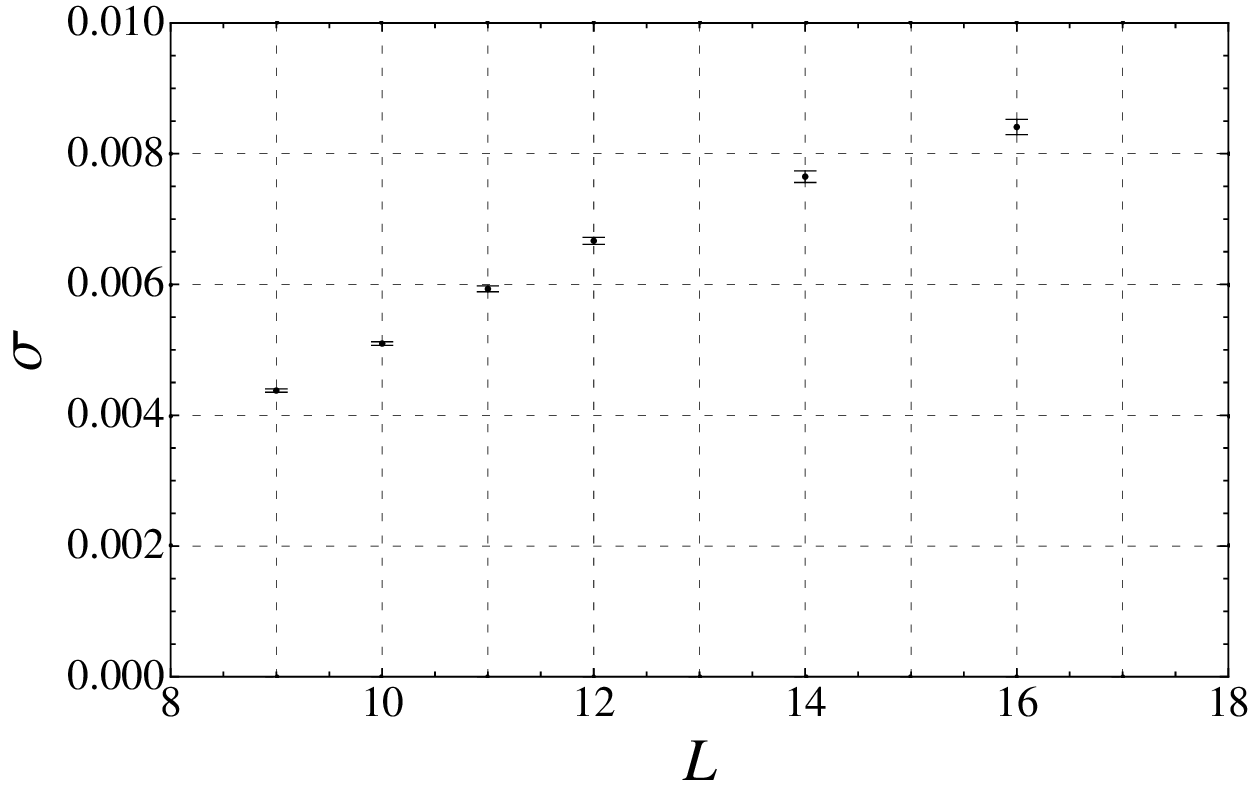}
  \caption{\small \textit{Plot of the string tension $\sigma$ extracted from $k(L)$ as a function of $L$.}}
  \label{fig:sigma_dependence}
\end{figure}

A few comments are in order on these fits:
\begin{description}
\item{1]} We used the following criteria to fix the range of values of $R$ used in the fits.
In order to fix the upper bound we performed a set of preliminary fits for the $L=10$ data keeping 
initially all the data and then iteratively discarding the largest ones looking for an acceptable $\chi^2_r$
(i.e. a reduced $\chi^2$ of order unity). In this way we identified as upper bound $R=50$.
For the lower bound we used the same criterion adopted in previous works on effective string corrections which
assumed the effective string to be a reliable description of the interquark potential
for scales $R$ such that $\sigma R^2 \geq 1$. In our case this means $R\geq 10$. Looking at (\ref{risfinale})
we see that with this choice the first subleading correction $\exp{-2\pi\frac{R}{L}}$ turns out to be negligible
within the statistical errors for all the values of $R$ and $L$ involved in the fits.

\item{2]} The linear fits show a reduced $\chi^2_r$ of order unity for all the values $L>9$
Also the data at $L=9$ show a linear like behaviour (see Fig. \ref{fig:simulation_results_08}) which is however
shadowed by rather large fluctuations. It is difficult to decide if this is the signature of a failure of the
effective string picture or if the fluctuations are simply due to vicinity of the critical temperature. In any
case we decided to neglect the $L=9$ data in the subsequent steps of our study.
 
\item{3]} We can compare the values of $c(L)$ extracted from the fits with the effective string
prediction:
\eq
  c(L)=\frac{1}{2\pi\sigma} \log\frac{L}{L_c}
\en

We  fitted the data for $L\geq10$ with the law:
$$
   c(L)=a \log{L} + b
$$

finding $a=17(4)$  and $b= -38(9)$ with a rather good value of
$\chi_r^2=1.6$. This result turns out to be in remarkable agreement with the effective string prediction:
 $a=1/2\pi\sigma = 15.12...$ (obtained assuming $\sigma=0.0105241$).

\item{4]}
Looking at Tab. \ref{tab:simulation_fit_08} we see that the values of $k(L)$ show a $L$ dependence different
from the one predicted by the effective string. However the values that we obtain from the fits smoothly converge
toward the predicted one as $L$ increase. This can be appreciated looking at the fourth column of 
Tab. \ref{tab:simulation_fit_08} where we reported the values of $k_0$ extracted from the fits assuming
$\sigma=0.0105241$ which should be
compared with the effective string prediction $k_0=1/4$. For future utility we also reported in the fifth column
of the table the values which we would predict for the string tension if we would fix $k_0=1/4$ in the fits.
These values are plotted in Fig. \ref{fig:sigma_dependence}.

\end{description}

\subsection{Discussion}
The above analysis shows that our data are in substantial agreement with the
effective string predictions at large values of $R$ and $L$. 
At the same time however we see an increasing disagreement as $R$ and $L$
decrease. This is indeed an usual phenomenon when effective string predictions
are compared  with LGT data and points to the well known fact that the effective string only represents
a large distance effective description of the interquark potential.
 
\begin{figure}[htpb]
  \centering
  \includegraphics[width=4 in]{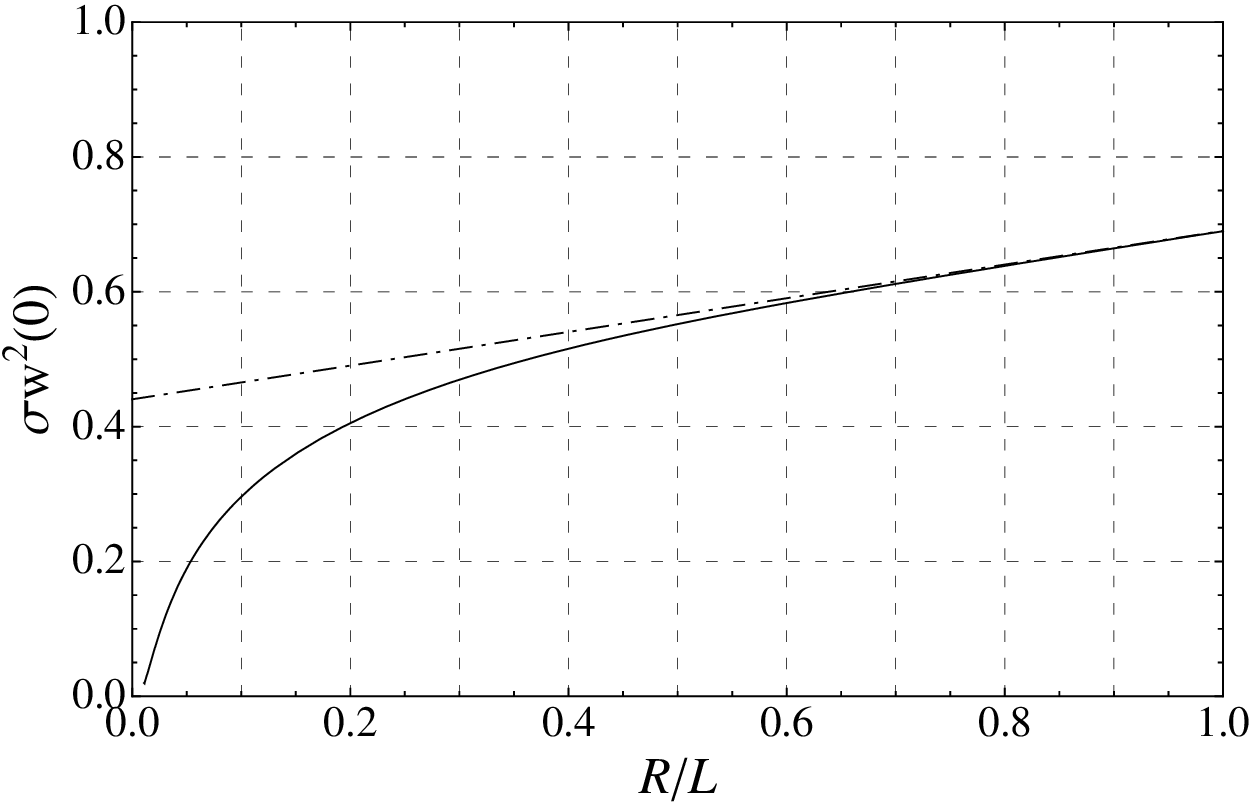}
  \caption{\small \textit{
 $R$ dependence of the flux tube width in the $R<L$ region. We also plot for comparison a 
 straight line with angular
 coefficient $1/4$.}}
  \label{fig:sqw_vs_R}
\end{figure}

We see two main sources of disagreement. 
The first one, which we already discussed above, is the 
fact that the effective string correction $1/4$ in front of the $k(L)$ term is only reached
asymptotically (see Fig. \ref{fig:sigma_dependence}). The second one is that looking at the data in Tab. \ref{res1} and Tab. \ref{res2} we see that the linear behaviour as a function of $R$ extends also in the $R<10$ region. This definitely disagrees with the effective string prediction which in this region (due to the higher terms in the theta functions) predicts instead a significant depletion of the flux tube width (see Fig. \ref{fig:sqw_vs_R}).
 
 It is interesting to observe that both these behaviours 
 instead agree with a naive Svetitsky-Yaffe~\cite{sy82} dimensional reduction picture (see.~\cite{Caselle:2006wr}).
 Indeed, according to this scheme, the correlation function of two Polyakov loops should behave as the spin-spin
 correlator of the 2d Ising model and the combination of plaquette and Polyakov loops used to measure the
 flux tube thickness is mapped into the $<\epsilon \sigma \sigma>$ correlator of the same Ising model.
 As discussed in~\cite{Caselle:2006wr} also in this framework one  finds a linear increase of the (equivalent of
 the) flux tube width, which however in this dimensional reduction scheme holds for all values of $R$. 
 Moreover the coefficient of this linear 
 increase is proportional to $1/m$ where $m$ is the mass of the equivalent 2d Ising model and depends on $L$ in a
 way which strongly resembles the behaviour reported in Fig. \ref{fig:sigma_dependence}
 (see~\cite{Caselle:2006wr}). 
  
 It would be interesting to see if going beyond the gaussian approximation one could 
 recover this behaviour also in the effective string framework. This would require  extending
 the calculation reported in the appendix to higher orders of the Nambu-Goto action or alternatively to adapt to
 this problem the D-brane approach discussed in~\cite{bc05} to deal with the whole Nambu-Goto action.  
 Work is in progress to test the feasibility of these approaches.

\newpage

\newpage
\appendix{}
\label{app}
\vskip 0.5cm
\section{{{Appendix A: Derivation of eq.(\ref{ris1},\ref{ris2})}}}
\renewcommand{\theequation}{A.\arabic{equation}}
\setcounter{equation}{0}
\subsection{The Green function for mixed boundary conditions}

The starting point of our analysis is the Green function for a free bosonic theory on a cylindric domain. This
Green function is the solution of the Laplace equation in a 2d rectangle with periodic boundary conditions in one
direction and Dirichlet conditions in the other direction. This can be mapped to to an electrostatic problem and
solved by the method of images.
We report the result in this appendix for completeness and discuss in some detail a few of its properties which
will later be relevant for the study of the flux tube thickness.

Let us map the cylinder onto a rectangle of the complex plane centered in the origin with sizes
 $[-L_x/2,L_x/2]\times[-L_y/2,L_y/2]$.  Let us impose periodic boundary conditions along the imaginary axis, i.e.
for $\Im(z)=\pm L_y/2$ and Dirichlet b.c. along the real axis i.e. for
$\Re(z)=\pm L_x/2$. 

Then the Green function can be written as:
\eq
G(z;z_0)=-\frac{1}{2 \pi}\log\left|f(z,z_0)
\right|
\label{app1}
\en
with:
\eq
\begin{array}{c}
\displaystyle
f(z;z_0)=
  \frac{\theta_1\left[ \pi (z-z_0)/2L_x\right]}
       {\theta_2\left[ \pi (z+\bar z_0)/2L_x\right]}
\end{array}
\label{app2}
\en
where the Jacobi theta functions $\theta_1$ and $\theta_2$ are defined as:
$$
\theta_1(z)=2q^{\frac14}\sum_{n=0}^{\infty}(-1)^n q^{n(n+1)} \sin(2n+1)z
$$

$$
\theta_2(z)=2q^{\frac14}\sum_{n=0}^{\infty} q^{n(n+1)} \cos(2n+1)z
$$
with:
$$
q= e^{i\pi \tau};\quad \tau=iL_y/2L_x
$$

In fact, with this definition, $\log f$ is analytic everywhere in the rectangle except in
 $z=z_0$ where $f(z)=0$ and the
Green function diverges logarithmically. For all the remaining values  since 
$\log |f|= \Re \log f$ (i.e. it is the real part of an analytic function) it
satisfies the Laplace equation:
$$
\Delta  G(z,z_0)=0
$$

As for the boundary conditions,
by using the transformation properties of the theta functions:
$$
  \theta_1(z+\pi\tau)=-q^{-1}e^{-2iz}\theta_1(z); \quad
  \theta_2(z+\pi\tau)=q^{-1}e^{-2iz}\theta_2(z);
$$
one can immediately see that:
$$G(z+iL_y,z_0)=G(z,z_0)$$
i.e. that $G$ is periodic along the imaginary axis with period $iL_y$

In order to check that eq (\ref{app1},\ref{app2}) satisfy the Dirichlet conditions along the $x$ axis, let us set
 $z=L_x/2+i y$. With this choice we have:
\[
G(z;z_0)=-\frac{1}{2 \pi}\log\left|
  \frac{\theta_1\left[ \pi (iy-z_0)/2L_x+\pi/4\right]}
       {\theta_2\left[ \pi (iy+\bar z_0)/2L_x+\pi/4\right]}
\right|
\]
from which, using the identity $\theta_2(z)=\theta_1(\pi/2-z)$ we immediately obtain:
$$
G(z;z_0)=-\frac{1}{2 \pi}\log\left|
  \frac{\theta_1\left[ \pi (iy-z_0)/2L_x+\pi/4\right]}
       {\theta_1\left[ \pi (-iy-\bar z_0)/2L_x+\pi/4\right]}
\right|
$$
Since the theta functions are real along the real axis, the denominator in the
above equation is the complex conjugate of the numerator and the argument of the logarithm is 
always unity for any value of $y$. In a similar way one can show that the Dirichlet b.c. 
hold also for $\Re z=-L_x/2$.
The expansion of eq.s (\ref{app1},\ref{app2}) converges very quickly when $L_y>>L_x$ . 
In this limit the exponentially
decreasing terms in $G(z,z_0)$ can be neglected and only the dominant terms in the theta functions give a
contribution. However in the opposite limit $L_x >> L_y$ such expression is almost useless. In this regime one
should better perform a modular transformation $\tau \to -1/\tau$ of the above result. 
In this way one obtains:

\eq
\begin{array}{c}
\displaystyle
G(z;z_0)=-\frac{1}{2 \pi}\log\left|
  \frac{\theta_1\left[i \pi (z-z_0)/L_y\right]}
       {\theta_4\left[i \pi (z+\bar z_0)/L_y\right]}
\right|
  +\frac{\mathrm{Re}z\ \mathrm{Re}z_0}{L_xL_y}\\ \\
\displaystyle
\quad q= e^{-2 \pi L_x/L_y};\quad \tau=2iL_x/L_y
\end{array}
\label{app4}
\en

This expression is equivalent to the above one, but converges well in the $L_x>>L_y$ limit.

\subsection{The flux tube width}

The behaviour of the flux tube width can be extracted from eq. (\ref{w4}) performing an expansion in $\epsilon$.
Keeping into account the prefactor $\sigma$ in the gaussian effective action eq. (\ref{SexpansionNLO}) 
we see that eq. (\ref{w4}) can be rewritten as: 

\eq
  \sigma w^2(x,y)= G(z,z+\epsilon)
\label{app5}
\en

In the limit $\epsilon\to0$, using eq.s (\ref{app1},\ref{app2}) and setting $L_x=R$, $L_y=L$ we obtain:

\eq
\begin{array}{c}
\displaystyle
\sigma w^2(z)=-\frac{1}{2\pi} \log\frac{\pi |\epsilon|}{2R} + 
\frac{1}{2\pi} \log\left|\theta_2\left(\pi \Re z/R\right)/\theta_1' \right|\\ \\
q=e^{-\pi L/2R} 
\end{array}
\label{app6}
\en

This expression converges well when $L>> R$. 

In the opposite regime $R >> L$ we must use eq. (\ref{app4})
which leads to:
\eq
\begin{array}{c}
  \displaystyle
  \sigma w^2(z)=-\frac{1}{2\pi} \log\frac{\pi |\epsilon|}{L} + \frac{1}{2\pi} \log\left|\theta_4(2\pi i \Re
  z/L)/\theta_1' \right| +
  \frac{(\Re z)^2}{LR}\\ \\
  q=e^{-\pi 2R/L}
\end{array}
  \label{app7}
\en

\vskip1.0cm {\bf Acknowledgements.}
 The authors would like to thank P.Grinza,
 F. Gliozzi, M. Bill\'o and L. Ferro for 
 useful discussions.

\end{document}